\documentclass[showpacs,showkeys,preprintnumbers, amsmath,amssymb,preprint]{revtex4-1}
\usepackage{amssymb}
\usepackage{amsfonts}
\usepackage{amsmath}
\usepackage{graphicx}
\usepackage{dcolumn}
\usepackage{bm}
\usepackage{makeidx}
\usepackage[latin1]{inputenc}
\usepackage[english, english]{babel}
\usepackage[dvips]{hyperref}

\usepackage{graphicx}
\def\be{\begin{equation}}
 \def\ee{\end{equation}}
 \def\bea{\begin{eqnarray}}
 \def\eea{\end{eqnarray}}

\makeindex

\begin{document}

\title{Scalar Perturbations 
of Nonlinear Charged Lifshitz Black Branes with Hyperscaling Violation}
\author{P. A. Gonz\'{a}lez}
\email{pablo.gonzalez@udp.cl}
\affiliation{Facultad de Ingenier\'{\i}a, Universidad Diego Portales, Avenida Ej\'{e}%
rcito Libertador 441, Casilla 298-V, Santiago, Chile.}
\author{Yerko V\'{a}squez.}
\email{yvasquez@userena.cl}
\affiliation{Departamento de F\'{\i}sica, Facultad de Ciencias, Universidad de La Serena,\\
Avenida Cisternas 1200, La Serena, Chile.}
\date{\today}

\begin{abstract}
We study scalar perturbations of nonlinear charged Lifshitz black branes with hyperscaling violating factor, and we find numerically the quasinormal modes for scalar fields. Then, we study the stability of these black branes under massive and massless scalar field perturbations.  Also, we consider different values of the dynamical exponent, the nonlinear exponent and the hyperscaling violating exponent.



\end{abstract}

\maketitle


\tableofcontents


\section{Introduction}

Lifshitz spacetimes have received considerable attention from the condensed matter point of view due to the AdS/CFT correspondence, i.e., the searching for gravity duals of Lifshitz fixed points for condensed matter physics and quantum chromodynamics \cite{Kachru:2008yh}. From the quantum field theory point of view, there are many invariant scale theories of interest
when studying such critical points. Such theories exhibit the anisotropic
scale invariance $t\rightarrow \chi ^{z}t$, $x\rightarrow \chi x$, with $z\neq
1$, where $z$ is the relative scale dimension of time and space, and these
are of particular interest in studies of critical exponent theory and phase
transitions. Systems with such behavior appear, for instance, in the
description of strongly correlated electrons. The importance of possessing a
tool to study strongly correlated condensed matter systems is beyond
question, and consequently much attention has focused on this area in
recent years. In this sense, Lifshitz holographic superconductivity has been a topic of numerous studies and interesting properties are found when one generalizes the gauge/gravity duality to non-relativistic situations \cite{Hartnoll:2009ns, Brynjolfsson:2009ct,  Sin:2009wi, Schaposnik:2012cr, Momeni:2012tw, Bu:2012zzb, Keranen:2012mx, Zhao:2013pva, Lu:2013tza, Tallarita:2014bga}.

The Lifshitz spacetimes are described by the metrics 
\begin{equation}
ds^2=- \frac{r^{2z}}{\ell^{2z}}dt^2+\frac{\ell^2}{r^2}dr^2+\frac{r^2}{\ell^2} d\vec{x}%
^2~,  \label{lif1}
\end{equation}
where $\vec{x}$ represents a $D-2$ dimensional spatial vector, $D$ is the
spacetime dimension and $\ell$ denotes the length scale in this geometry. If $z=1$, the spacetime is the
usual anti-de Sitter metric in Poincar\'{e} coordinates. Furthermore, all scalar
curvature invariants are constant and these spacetimes have a null curvature
singularity at $r\rightarrow 0$ for $z\neq 1$, which can be seen by
computing the tidal forces between infalling particles. This singularity is
reached in finite proper time by infalling observers, so the spacetime is
geodesically incomplete \cite{Horowitz:2011gh}. The metrics of Lifshitz
black holes asymptotically have the form (\ref{lif1}); however, obtaining
analytical solutions does not seem to be a trivial task, and therefore
constructing finite temperature gravity duals requires the introduction of
strange matter content with a theoretical motivation that is not clear.
Another way of finding such a Lifshitz black hole solution is by considering
carefully-tuned higher-curvature modifications to the Hilbert-Einstein
action, as in new massive gravity (NMG) in 3-dimensions or $R^2$ corrections
to general relativity. Some Lifshitz black holes solutions have been found in references \cite{Balasubramanian:2009rx, Mann:2009yx, AyonBeato:2009nh, Bertoldi:2009vn, Cai:2009ac, AyonBeato:2010tm, Dehghani:2010kd}.
Thermodynamically, it is difficult to compute conserved
quantities for Lifshitz black holes; however, progress was made on the
computation of mass and related thermodynamic quantities by using the ADT
method \cite{Devecioglu:2010sf, Devecioglu:2011yi} and the off-shell extension of the ADT formalism \cite{Gim:2014nba} as well as the Euclidean action approach \cite{Gonzalez:2011nz, Myung:2012cb}. Also, phase
transitions between Lifshitz black holes and other configurations with
different asymptotes have been studied in \cite{Myung:2012xc}. However, due
to their different asymptotes these phases transitions do not occur. 

A generalization of the above metric is given by
\begin{equation}
ds^2= r^{-\frac{2 \theta}{D}} \left(- \frac{r^{2z}}{\ell^{2z}}dt^2+\frac{\ell^2}{r^2}dr^2+\frac{r^2}{\ell^2} d\vec{x}%
^2 \right)~,
\end{equation}
which, besides having  an anisotropic scaling as the Lifshitz metric, have an overall hyperscaling violating factor with hyperscaling exponent $\theta$, thus, this line element is conformally related to the Lifshitz metric. This space-time is important in the study of the dual field theories with hyperscaling violation \cite{Dong:2012se, Gath:2012pg}. Lifshitz black holes with hyperscaling violation have been found in \cite{Dehghani:2015gza, Feng:2015yja, Ganjali:2015cba}.


In this work, we study scalar perturbations of nonlinear charged Lifshitz black branes with hyperscaling violation.
The matter is parameterized by scalar fields minimally coupled to gravity. Then, we obtain numerically the quasinormal frequencies (QNFs) for scalar fields. We focus our study in the influence of the dynamical  exponent, the nonlinear exponent and the hyperscaling exponent in the stability. 




The study of the QNFs \cite{Regge:1957td, Zerilli:1971wd,
Zerilli:1970se, Kokkotas:1999bd, Nollert:1999ji, Konoplya:2011qq} gives
information about the stability of black holes under matter fields that
evolve perturbatively in their exterior region, without backreacting on the
metric. In general, the oscillation frequencies are complex, where the real
part represents the oscillation frequency and the imaginary part describes
the rate at which this oscillation is damped, with the stability of the
black hole being guaranteed if the imaginary part is negative. The
QNFs are independent of the initial
conditions and depend only on the parameters of the black hole (mass,
charge and angular momentum) and the fundamental constants (Newton constant
and cosmological constant) that describe a black hole, just like the
parameters that define the test field.  On the other hand, the QNFs determine how fast a thermal state in the
boundary theory will reach thermal equilibrium according to the AdS/CFT
correspondence \cite{Maldacena:1997re}, where the relaxation time of a
thermal state is proportional to the inverse
of the imaginary part of the QNFs of the dual gravity background, which was
established due to the QNFs of the black hole being related to the poles of
the retarded correlation function of the corresponding perturbations of the
dual conformal field theory \cite{Birmingham:2001pj}. Fermions on a Lifshitz
background were studied in \cite{Alishahiha:2012nm} by using the
fermionic Green's function in 4-dimensional Lifshitz spacetime with $z=2$; the authors considered a non-relativistic (mixed) boundary condition
for fermions and showed that the spectrum has a flat band. Also, the Dirac quasinormal modes (QNMs) for a 4-dimensional Lifshitz black hole were studied in \cite{Catalan:2013eza}. Generally, the Lifshitz black holes are stable under scalar perturbations, quasinormal modes under scalar field perturbations have been studied in \cite{CuadrosMelgar:2011up, Gonzalez:2012de, Gonzalez:2012xc, Myung:2012cb, Becar:2012bj,Giacomini:2012hg, Lepe:2012zf, Catalan:2014ama, Catalan:2014una} and electromagnetic quasinormal modes in \cite{Lopez-Ortega:2014oha}. Moreover, it is was stablished that for $d> z+1$ , at zero momenta, the modes are non-overdamped, whereas for $d \leq z + 1$ the system is always overdamped \cite{Sybesma:2015oha}. The QNFs have been calculated by means of numerical and analytical techniques, some remarkably numerical methods are: the Mashhoon method, Chandrasekhar-Detweiler, WKB method, Frobenius method, method of continued fractions, Nollert, asymptotic iteration method (AIM) and improved AIM among others. In the context of black hole
thermodynamics, QNMs allow the quantum area spectrum of the black hole
horizon to be studied as well as the mass and
the entropy spectrum.


The paper is organized as follows. In Sec. \ref{Background} we give a brief review
 of nonlinear charged Lifshitz black branes with hyperscaling violation that we will consider as background. In Sec. \ref{QNM}
we calculate the QNFs of scalar perturbations numerically by using the improved AIM. Finally, our conclusions are in Sec. \ref{conclusion}.

\section{Nonlinear charged Lifshitz black branes with hyperscaling violation}
\label{Background}
The nonlinear charged Lifshitz black brane that we consider is a solution of the Einstein-dilaton gravity in the presence of a linear and a nonlinear electromagnetic field, this solution was found in \cite{Dehghani:2015gza}. The action is given by
\begin{equation}\label{action4d}
S=\frac{1}{16\pi}\int_{\mathcal{M}} d^{D}x\sqrt{-g}\left( R -\frac{1}{2}(\nabla\phi)^2+V(\phi)-\frac{1}{4}e^{\lambda_{1}\phi}H_{\mu \nu}H^{\mu \nu}+\frac{1}{4}e^{\lambda_{2}\phi}(-F)^s\right)\,,
\end{equation}
where $R$ is the Ricci scalar on the manifold $\mathcal{M}$, $\phi$ is the dilatonic field, $\lambda_{1}$ and $\lambda_{2}$ are free parameters of the model, $F$ is the Maxwell invariant of the electromagnetic field $F_{\mu\nu}=\partial_{[\mu}A_{\nu]}$, where $A_\mu$ is the electromagnetic potential and $H_{\mu\nu}=\partial_{[\mu}B_{\nu]}$ is a linear electromagnetic field, where $B_\mu$ is the electromagnetic potential.
The following metric is solution of the theory defined by the action (\ref{action4d}), and it represents a black brane solution with hyperscaling violating factor
\begin{equation}\label{metric}
ds^2=r^{2 \alpha} \left(-r^{2z}f(r)dt^2+\frac{dr^2}{r^2 f(r)}+r^2 \sum _{i=1}^{D-2} dx_{i}^2 \right)\,,
\end{equation}
with
\begin{equation}
f(r)=1-\frac{M}{r^{z+D-2-\theta}}+\frac{Q^{2s}}{r^{\Gamma+D-2+z-\theta}}\,,
\end{equation}
where $\alpha=-\frac{\theta}{D-2}$ has been used, and $\theta$ is the hyperscaling exponent, also
\begin{eqnarray}
\notag  &&Q^{2s}=\frac{(2s-1)r_{0}^{2(z-1-\frac{\theta}{D-2})}}{4(D-2-\theta)\Gamma}(2q_2^2)^s\,, \\
 &&\Gamma=z-2+\frac{D-2-\theta}{2s-1}\,,
\end{eqnarray}
where $M$ is an integration constant related to the mass of the black brane and $q_{2}$ is an integration constant related to its electric charge. The solutions are not valid for $\alpha=-1$. To have $f(r) \rightarrow 1$ when $r \rightarrow \infty$, the following inequalities must be satisfied
\begin{equation}\label{ineq}
z+D-2-\theta>0~, \,\,\, \Gamma+D-2+z-\theta>0\,,
\end{equation}
however, it can be shown that $\Gamma>0$ \cite{Dehghani:2015gza}.
The gauge and dilatonic fields are given by
\begin{eqnarray}
\notag &&F_{rt}=q_{2}r_{0}^{2\left( -\frac{\theta}{D-2}+z-1 \right)}r^{-\left( \Gamma+1 \right)}\,, \\
\notag &&H_{rt}=q_{1}r_{0}^{2\left( -\frac{\theta}{D-2}+\theta-1 \right)}r^{\left( D-2+z-1 \right)}\,, \\
&&\phi (r)=ln \left( \frac{r}{r_{0}}\right)^{\sqrt{2(D-2-\theta)\left(-\frac{\theta}{D-2}+z-1 \right)}}\,,
\end{eqnarray}
thus, for a real dilatonic field we must have $(D-2-\theta)\left(-\frac{\theta}{D-2}+z-1 \right) \geq 0$. Moreover, since $z+D-2-\theta>0$, $z$ has to be larger than $1$ too. It is worth to mention that the condition for having a black hole is
\begin{equation}
\left( \frac{\Gamma M}{\Gamma +z+D-2-\theta} \right)^{\Gamma +z+D-2-\theta} \geq \left( \frac{\Gamma Q^{2s}}{z+D-2-\theta} \right)^{z+D-2-\theta}\,.
\end{equation}
The temperature and entropy of the solution are given by
\begin{equation}
T=\frac{1}{4 \pi}\left((z+D-2-\theta)r_{h}^z-\Gamma Q^{2s} r_{h}^{-(\Gamma+D-2-\theta)} \right)\,,\,\,\,\, S=\frac{1}{4}r_{h}^{D-2-\theta}\,,
\end{equation}
where $r_{h}$ denotes the event horizon. The study of the thermodynamics was performed in detail in \cite{Dehghani:2015gza}.


 
\section{Quasinormal modes}
\label{QNM}
The QNMs of scalar perturbations in the background of a $D$-dimensional nonlinear charged Lifshitz black brane are given by the scalar field solution of the Klein-Gordon equation with suitable boundary conditions for a black brane geometry. This means there are only ingoing waves on the event horizon and we consider that the scalar field vanishes at spatial infinity, known as Dirichlet boundary condition. 
The Klein-Gordon equation for a scalar field minimally coupled to curvature is 
\begin{equation}
\frac{1}{\sqrt{-g}}\partial _{\mu }\left( \sqrt{-g}g^{\mu \nu }\partial
_{\nu } \varphi \right) =m^{2}\varphi \,,  \label{KGNM}
\end{equation}%
where $m$ is the mass of the scalar field $\varphi $. 
Now, by means of the following ansatz 
\begin{equation}
\varphi =e^{-i\omega t}e^{i\vec{\kappa} \cdot \vec{x}}R(r)\,,
\end{equation}%
where $\vec{x}$ is a spatial vector in $D-2$ dimensions, and $-\kappa^2$ is the eigenvalue of the Laplacian in the flat base submanifold.
The Klein-Gordon equation reduces to
\begin{equation}
 \frac{1}{r^{\beta}}\frac{d}{dr}\left(r^{2+\beta-2\alpha}f(r)\frac{dR}{dr}\right)+\left(\frac{\omega^{2}}{r^{2\alpha+2z}f(r)}-\frac{\kappa ^2}{r^{2\alpha+2}}-m^{2}\right) R(r)=0\,, \label{radial}
\end{equation}%
where $\beta=(\alpha+1)D+z-3$ has been defined. Now, defining $R(r)$ as 
 \begin{equation}
 R(r)=\frac{F(r)}{r^{n}}\,,
 \end{equation}
where $n=\frac{(D-2)(1+\alpha)}{2}$, and by using the tortoise coordinate $x$ given by  
 \begin{equation}
 dx=\frac{dr}{r^{z+1}f(r)}\,,
 \end{equation}
 the Klein-Gordon equation can be written as a one-dimensional Schr\"{o}dinger equation 
 \begin{equation}\label{ggg}
 \frac{d^{2}F(x)}{dx^2}-V(r)F(x)=-\omega^{2}F(x)\,,
 \end{equation}
 with an effective potential $V(r)$ given by
 \begin{equation}
 V(r)=\frac{1}{4}r^{2z-2}f(r)\left(4(m^2 r^{2+2\alpha}+\kappa^2)+(D-2)(1+\alpha)\left( \left( (D-2)(1+\alpha)+2z \right) f(r)+2r\frac{df}{dr} \right)r^2\right)\,,
 \end{equation}
that diverges at spatial infinity.
It is worth to mention that is not trivial to find analytical solutions to Eq. (\ref{radial}). 
In the next section, we will perform numerical studies 
by using the improved AIM \cite{Cho:2009cj}, which is an improved version of the method proposed in references \cite{Ciftci, Ciftci:2005xn} and it has been applied successful in the context of QNMs for different black holes geometries, see for instance \cite{Cho:2009cj, Cho:2011sf, Catalan:2013eza, Zhang:2015jda, Barakat:2006ki}.


\subsubsection{Stability analysis}

Following the argument used in \cite{Horowitz:1999jd}, adapted to Lifshitz geometries with hyperscaling violation, we can verify when the imaginary part of the quasinormal frequency $\omega$ is always negative. By using outgoing Eddington-Filkenstein coordinates $v=t+x$, metric (\ref{metric}) can be transformed to
\begin{equation}
ds^2=r^{2\alpha}\left( r^{2z}f(r)dv^2+2r^{z-1}dvdr+r^2\sum _{i=1}^{D-2}dx_{i}^2 \right)\,.
\end{equation}
Now, taking as ansatz
\begin{equation}
\varphi=e^{-i\omega t}e^{i\vec{\kappa} \cdot \vec{x}} \frac{\psi(r)}{r^n}\,,
\end{equation}
with $n=\frac{(D-2)(1+\alpha)}{2}$, the Klein-Gordon equation yields
\begin{equation}\label{KleinFink}
\frac{d}{dr}(r^{1+z}f(r)\psi'(r))-2i\omega\psi'(r)-V(r)\psi(r)=0\,,
\end{equation}
where  
\begin{equation}\label{pot}
V(r)=nr^{z}f'(r)+n(n+z)r^{z-1}f(r)+\kappa^2r^{z-3}+m^2 r^{z+2\alpha-1}\,.
\end{equation}
Notice that $n>-\frac{z}{2}$ according to the inequalities (\ref{ineq}). Then, multiplying equation (\ref{KleinFink}) by $\psi^{\ast}$ and performing integrations by parts, and using Dirichlet boundary condition for the scalar field at spatial infinity, one can obtain
\begin{equation}\label{relacion}
\int _{r_{h}}^{\infty}dr \left( r^{1+z}f(r) \left|  \frac{d\psi}{dr}\right|^2+V(r) \left| \psi \right| ^2 \right)=-\frac{\left|\omega \right|^2 \left| \psi (r=r_{h})\right| ^2}{Im(\omega)}\,,
\end{equation}
thus, the stability of the black brane under scalar field perturbations is guaranteed for a strictly positive potential $V(r)$ outside the horizon, because in this case, equation (\ref{relacion}) is satisfied only for $Im(\omega)<0$. Notice that the potential (\ref{pot}) is positive for $n>0$ or $\alpha>-1$, what guaranties the stability of the black brane solution. In this work we focus our attention to this case ($\alpha>-1$).

\subsubsection{Numerical analysis}

In order to implement the improved AIM 
we make the following change of variable $y=1-r_{h}/r$ to equation (\ref{radial}), where $r_{h}$ denotes the location of the event horizon, thus, in this coordinate, the event horizon is located at $y=0$ and the spatial infinity at $y=1$. Then, the Klein-Gordon equation becomes
\begin{eqnarray}
\nonumber && \frac{d^2R}{dy^2}+\left(-\frac{2\alpha-\beta}{1-y}+\frac{f^{\prime}(y)}{f(y)}\right) \frac{dR}{dy}\\
&&+\left( \frac{\omega^2(1-y)^{2z-2}}{r_h^{2z}f(y)^2}-\frac{\kappa ^2}{r_h^2 f(y)}-\frac{m^2 r_{h}^{2 \alpha}}{(1-y)^{2\alpha+2} f(y)}\right)R=0\,,
\label{numericalmethod}
\end{eqnarray}
in this equation $f(y)$ refers to the function $f(r)$ evaluated at $r=\frac{r_{h}}{1-y}$; that is
\begin{equation}
f(y)=1-\frac{M (1-y)^{z+D-2-\theta}}{r_{h}^{z+D-2-\theta}}+\frac{Q^{2s}(1-y)^{\Gamma+D-2+z-\theta}}{r_{h}^{\Gamma+D-2+z-\theta}}\,,
\end{equation}
and $f'(y)=\frac{df(y)}{dy}$.
Now, we must consider the behavior of the scalar field on the event horizon and at spatial infinity. First, notice that the tortoise coordinate is given in terms of the $y$ coordinate by
\begin{equation}
dx=\frac{dr}{r^{z+1}f(r)}=\frac{(1-y)^{z-1}dy}{r_{h}^{z}f(y)}\,.
\end{equation}

\begin{itemize}

\item{Event horizon}

In the limit $y \rightarrow 0$ ($r \rightarrow r_{h}$) the function $f(y)$ tends to $f(y)=f'(y)y+\mathcal{O}(y^2)$, where $\mathcal{O}(y^2)$ denotes terms of order $y^2$ and higher which can be neglected when $y \rightarrow 0$. Therefore, the tortoise coordinate is given explicitly by
\begin{equation}
x=\frac{ln(y)}{f'(0)r_{h}^{z}}\,,
\end{equation}
where $f'(0)=\frac{df(y)}{dy}\vert _{y=0}$ and the effective potential $V(x)$ tends to zero in this limit; thus, equation (\ref{ggg}) reduces to
\begin{equation}
\frac{d^2F(x)}{dx^2}=-\omega^2F(x)\,,
\end{equation}
and its solution is
\begin{equation}
F(x)=C_{1}e^{-i\omega x}+C_{2}e^{i\omega x}\,.
\end{equation}
Imposing as boundary condition that only ingoing waves exist on the event horizon, we must set $C_{2}=0$. Therefore, the solution near the horizon is given by
\begin{equation}
F(x)=C_{1}e^{-i\omega x} \sim  y^{-\frac{i\omega}{r_{h}^{z}f'(0)}}\,.
\end{equation}

\item{Spatial infinity}

On the other hand, when $y \rightarrow 1$ ($r\rightarrow \infty$), the tortoise coordinate is given by
\begin{equation}
x=-\frac{(1-y)^z}{zr_{h}^z}\,.
\end{equation}
In the following we will consider two cases: the first case corresponds to $-1<\alpha<0$, and the second case corresponds to $\alpha>0$ and $m=0$.
In both cases the effective potential tends to:
\begin{equation}
V(x)=\frac{\delta}{z^2x^2}\,,
\end{equation}
where $\delta=n(n+z)=\frac{1}{4} (1+\alpha)(D-2)(2z+(D-2)(1+\alpha))$. Then, equation (\ref{ggg}) for $x \rightarrow 0$ becomes
\begin{equation}
\frac{d^2F(x)}{dx^2}-\frac{\delta}{z^2 x^2}F(x)=0\,,
\end{equation}
whose solution is
\begin{equation}
F(x)=D_{1}x^{\frac{1}{2}(1-\sqrt{1+\frac{4\delta}{z^2}})}+D_{2}x^{\frac{1}{2}(1+\sqrt{1+\frac{4\delta}{z^2}})}\,.
\end{equation}
Notice that the effective potential asymptotically tends to $+\infty$ or to $-\infty$ depending if $\delta$ is positive or negative, respectively. In this work we focus on $\delta >0$, and imposing Dirichlet boundary condition, that is, to have a null field at spatial infinity, we must set $D_{1}=0$. Therefore the solution becomes
\begin{equation}
F(x)=D_{2}x^{\frac{1}{2}(1+\sqrt{1+\frac{4\delta}{z^2}})}\sim(1-y)^{\frac{1}{2}z(1+\sqrt{1+\frac{4\delta}{z^2}})}\,.
\end{equation}
Notice that, due to (\ref{ineq}), $n+z>z/2$. So, requiring $\delta=n(n+z)>0$ implies $n>0$, or $\alpha>-1$.
\end{itemize}
Thus, taking into account these behaviors we define 
\begin{equation}
R\left( y\right) = y^{-\frac{i\omega}{r_{h}^{z}f'(0)}} (1-y)^{\frac{1}{2}z(1+\sqrt{1+\frac{4\delta}{z^2}})} \chi (y)\,,
\end{equation}
as ansatz. Then, by inserting these fields in Eq. (\ref{numericalmethod}) we obtain the homogeneous
linear second-order differential equation for the function $\chi (z)$ 
\begin{equation}
\chi ^{\prime \prime }=\lambda _{0}(y)\chi ^{\prime }+s_{0}(y)\chi \,,
\label{de}
\end{equation}%
where
\begin{eqnarray}
\notag \lambda _{0}(y)&=&\frac{1}{r_{h}^z f'(0)y(1-y)f(y)} (( r_{h}^z f'(0) y (3+2 \alpha-D(1+\alpha)+z \sqrt{1+4 \delta/z^2}) \\
&&+ 2i (1-y) \omega )f(y)-r_{h}^z f'(0) y(1-y) f'(y)) \,, \\
\notag s_{0}(y)&=& -\frac{1}{2r_{h}^{2(z+1)}f'(0)^2y^2(1-y)^{2(1+\alpha)}f(y)^2}(2r_{h}^2f'(0)^2 \omega^2 y^2(1-y)^{2(z+\alpha)}-r_{h}^2(1-y)^{2 \alpha} \\
\notag &&( r_{h}^{2z}f'(0)^2y^2 (-2\delta+(D-2)z(1+\alpha)(1+\sqrt{1+4\delta/z^2}))-2ir_{h}^zf'(0)(1-y) \\
\notag &&(1-(D-2)(1+\alpha)y+z \sqrt{1+4\delta/z^2}y)\omega+2(1-y)^2\omega^2)f(y)^2-r_{h}^zf'(0)yf(y) \\
\notag && (2r_{h}^zf'(0)y(r_{h}^{2 \alpha+2}m^2+\kappa^2 (1-y)^{2+2\alpha})+r_{h}^2 (1-y)^{1+2\alpha}(r_{h}^zf'(0)z(1+\sqrt{1+4\delta/z^2})y \\
&&+2i(1-y)\omega)f'(y)))\,.
\end{eqnarray}

Then, in order to implement the improved AIM it is necessary to differentiate Eq. (\ref{de}) $n$ times with respect to $y$,
which yields the following equation: 
\begin{equation}
\chi ^{n+2}=\lambda _{n}(y)\chi ^{\prime }+s_{n}(y)\chi~,  \label{de1}
\end{equation}%
where 
\begin{equation}
\lambda _{n}(y)=\lambda _{n-1}^{\prime }(y)+s_{n-1}(y)+\lambda
_{0}(y)\lambda _{n-1}(y)~,  \label{Ln}
\end{equation}%
\begin{equation}
s_{n}(y)=s_{n-1}^{\prime }(y)+s_{0}(y)\lambda _{n-1}(y)\,.  \label{Snn}
\end{equation}%
Then, by expanding the $\lambda _{n}$ and $s_{n}$ in a Taylor series around
some point $\eta $, at which the improved AIM is performed yields
\begin{equation}
\lambda _{n}(\eta )=\sum_{i=0}^{\infty }c_{n}^{i}(y-\eta )^{i}\,,
\end{equation}%
\begin{equation}
s_{n}(\eta )=\sum_{i=0}^{\infty }d_{n}^{i}(y-\eta )^{i}\,,
\end{equation}%
where the $c_{n}^{i}$ and $d_{n}^{i}$ are the $i^{th}$ Taylor coefficients
of $\lambda _{n}(\eta )$ and $s_{n}(\eta )$, respectively, and by replacing
the above expansions in Eqs. (\ref{Ln}) and (\ref{Snn}) the following
set of recursion relations for the coefficients is obtained:
\begin{equation}
c_{n}^{i}=(i+1)c_{n-1}^{i+1}+d_{n-1}^{i}+%
\sum_{k=0}^{i}c_{0}^{k}c_{n-1}^{i-k}~,
\end{equation}%
\begin{equation}
d_{n}^{i}=(i+1)d_{n-1}^{i+1}+\sum_{k=0}^{i}d_{0}^{k}c_{n-1}^{i-k}\,.
\end{equation}%
In this manner, the authors of the improved AIM have avoided the
derivatives that contain the AIM in \cite{Cho:2009cj, Cho:2011sf}, and
the quantization condition, which is equivalent to imposing a termination
to the number of iterations, is given by 
\begin{equation}
d_{n}^{0}c_{n-1}^{0}-d_{n-1}^{0}c_{n}^{0}=0~.
\end{equation}
We solve this equation numerically 
and we choose different values for the parameters. Thus, without loss of generality, we choose the following values $D=4$, $M=4$, $z=3$, $r_0=1$, $m=0$ and $\kappa=0$ in Tables \ref{QNM2}, \ref{QNM3} and \ref{QNM1ml}. In Table \ref{QNM2}, we show some lowest QNFs, for massless scalar field  with $s=1.5$, $q_2=1$ and different values of $\theta$. Then, in Table \ref{QNM3}, we show some lowest QNFs, for $\theta =1$, $q_2=1$ and different values of $s$ and in Table \ref{QNM1ml} we show QNFs with $\theta =1$, $s=2$ and different values of $q_2$. In Table \ref{QNM2ml} we show QNFs for $D=4$, $M=4$, $\theta =1$, $s=2$, $r_0=1$, $q_2=1$, $m=0$, $\kappa=0$ and different values of $z$. We observe that in all cases analyzed the QNFs have an imaginary part that is negative, which ensures the stability of nonlinear charged Lifshitz black branes with hyperscaling violation under massless scalar  perturbations. 
Then, in Table \ref{QNM1p}  we show some lowest QNFs for massive scalar fields, for $s=2$, $\theta=1$, $\kappa=1$, and different values of $m$. Finally, in Table \ref{QNM3ml} we show fundamental QNFs  for massless scalar field for $D=4$ and $D=5$, with different values of the dynamical and hyperscaling violating exponents. Note that the Klein-Gordon equation depends on the combination $D-\theta$, this is the reason why, for instance, the same QNFs are obtained for $D=4$, $\theta=0.5$ and $D=5$, $\theta=1.5$ in Table \ref{QNM3ml}.

\begin{table}[ht]
\caption{Some quasinormal frequencies for $D=4$, $M=4$, $s=1.5$, $r_0=1$, $q_2=1$, $z=3$, $m=0$, $\kappa=0$ and different values of $\theta$.}
\label{QNM2}\centering
\begin{tabular}{cccccccccc}
\hline\hline
$n$ & $\theta=-1$ & $\theta=-0.5$ & $\theta=0$ & $\theta=0.5$ & $\theta=1$ & $\theta=1.5$  \\[0.5ex] \hline
$0$ & $-10.71750i$ & $-8.77694i$ &  $-7.57904i$ &  $-6.57065i$ &  $-5.45750i$ &  $-2.85448i$ \\ 
$1$ & $-12.82510i$ & $-13.22240i$  & $-12.62600i$ &  $-11.65420i$ &  $-10.15690i$ &  $-5.56952i$  \\ 
$2$ & $-21.81980i$ & $-19.50890i$ & $-18.14820i$ &  $-16.81610i$ &  $-14.83420i$ &  $-8.26014i$ \\ 
$3$ & $-25.13070$ & $-24.87590i$ &  $-23.57970i$ &  $-21.99220i$ &  $-19.51670i$ &  $-10.94460i$ \\[0.5ex] \hline
\end{tabular}%
\end{table}

\begin{table}[ht]
\caption{Quasinormal frequencies for $D=4$, $M=4$, $z=3$, $\theta=1$, $r_0=1$, $q_2=1$, $m=0$, $\kappa=0$ and different values of $s$.}
\label{QNM3}\centering
\begin{tabular}{cccccc}
\hline\hline
$n$ & $s=1.5$ & $s=2$ & $s=2.1$ \\[0.5ex] \hline
$0$ & $-5.45750i$ &  $-3.16787i$ & $-2.22508i$ \\ 
$1$ & $-10.15690i$ &   $-6.05288i$ & $-4.30905i$ \\ 
$2$ & $-14.83420i$ &   $-8.87462i$ & $-6.34705i$ \\ 
$3$ & $-19.51670i$ &   $-11.68160i$ & $-8.36704i$ \\[0.5ex] \hline
\end{tabular}%
\end{table}
\begin{table}[ht]
\caption{Quasinormal frequencies for $D=4$, $M=4$, $z=3$, $s=2$, $\theta=1$, $r_0=1$ $m=0$, $\kappa=0$ and different values of $q_2$.}
\label{QNM1ml}\centering
\begin{tabular}{cccccc}
\hline\hline
$n$ & $q_2=0.1$ & $q_2=0.5$ & $q_2=1$ \\[0.5ex] \hline
$0$ & $-6.66991i$ & $-6.49942i$ & $-3.16787i$ \\ 
$1$ & $-12.25930i$ & $-11.96360i$ & $-6.05288i$ \\ 
$2$ & $-17.91470i$ & $-17.48060i$ & $-8.87462i$ \\ 
$3$ & $-23.57020i$ & $-22.99920i$ & $-11.68160i$ \\[0.5ex] \hline
\end{tabular}%
\end{table}
\begin{table}[ht]
\caption{Quasinormal frequencies for $D=4$, $M=4$, $\theta=1$, $s=2$, $r_0=1$, $q_2=1$, $m=0$, $\kappa=0$ and different values of $z$.}
\label{QNM2ml}\centering
\begin{tabular}{cccccc}
\hline\hline
$n$ & $z=3$ & $z=5$ & $z=8$ \\[0.5ex] \hline
$0$ & $-3.16787i$ &  $-8.80358i$ & $-15.17150i$ \\ 
$1$ & $-6.05288i$ &  $-16.87330i$ & $-29.48100i$ \\ 
$2$ & $-8.87462i$ &  $-24.90450i$ & $-43.77070i$ \\ 
$3$ & $-11.68160i$ &  $-32.93630i$ & $-58.06180i$ \\[0.5ex] \hline
\end{tabular}%
\end{table}
\begin{table}[ht]
\caption{Quasinormal frequencies for $D=4$, $M=4$, $z=3$, $\theta=1$, $r_0=1$, $q_2=1$, $\kappa=1$, $s=2$ and different values of $m$.}
\label{QNM1p}\centering
\begin{tabular}{cccccc}
\hline\hline
$n$ & $m=1$ & $m=2$ & $m=3$ \\[0.5ex] \hline
$0$ & $-3.49027i$ &  $-3.94694i$ & $-4.58214i$ \\ 
$1$ & $-6.32653i$ &  $-6.75628i$ & $-7.37113i$ \\ 
$2$ & $-9.11635i$ &  $-9.52451i$ & $-10.12130i$ \\ 
$3$ & $-11.89980i$ &  $-12.28760i$ & $-12.86430i$ \\[0.5ex] \hline
\end{tabular}%
\end{table}
\begin{table}[ht]
\caption{Fundamental quasinormal frequencies for $M=4$, $Q=1$, $m=0$, $\kappa=0$, $s=2$, $r_0=1$, $q_2=1$ and different values of $D$, $z$ and $\theta$. (In the empty rows no event horizon exist or $\Gamma<0$).}
\label{QNM3ml}\centering
\begin{tabular}{ | r | r | r | r | r | r |  }
\hline
\multicolumn{6}{|c|}{D=4} \\ \hline
$z$ & $\theta=-1$ & $\theta=0$ & $\theta=0.5$ & $\theta=1$ & $\theta=1.5$ \\ \hline
$1.5$ & $2.86396-5.49737i$ &  $--$ & $--$ & $--$ &  $--$ \\ 
$2$ & $2.71457-7.93528i$ &  $-3.50142i$ & $--$ & $--$ &  $--$ \\ 
$2.5$ & $1.54271-9.78902i$ &  $-5.52189i$ & $-3.64301i$ & $--$ &  $--$ \\ 
$3$ & $-9.75248i$ &  $-6.70287i$ & $-5.34727i$ & $-3.16787i$ &  $--$ \\ 
$3.5$ & $-10.11190i$ &  $-7.73738i$ & $-6.61238i$ & $-5.01956i$ &  $--$ \\ 
$4$ & $-10.76490i$ &  $-8.72993i$ & $-7.74275i$ & $-6.41728i$ &  $-2.33210i$ \\ 
$4.5$ & $-11.52580i$ &  $-9.70835i$ & $-8.81550i$ & $-7.65052i$ &  $-4.61700i$ \\ \hline
\multicolumn{6}{|c|}{D=5} \\ \hline
$z$ & $\theta=-1$ & $\theta=0$ & $\theta=0.5$ & $\theta=1$ & $\theta=1.5$ \\ \hline
$1.5$ & $5.45477-5.84803i$ & $2.86396-5.49737i$ &  $--$ & $--$ & $--$ \\
$2$ & $5.38082-8.28982i$ & $2.71457-7.93528i$ &  $0.25590-7.06029i$ & $-3.50142i$ & $--$ \\
$2.5$ & $4.76150-10.52210i$ & $1.54271-9.78902i$ &  $-7.28570i$ & $-5.52189i$ & $-3.64301i$ \\
$3$ & $3.74054-12.47720i$ & $-9.75248i$ &  $-8.01991i$ & $-6.70287i$ & $-5.34727i$ \\
$3.5$ & $2.29788-14.15880i$ & $-10.11190i$ &  $-8.83530i$ & $-7.73738i$ & $-6.61238i$ \\
$4$ & $-14.10110i$ & $-10.76490i$ &  $-9.69375i$ & $-8.72993i$ & $-7.74275i$ \\
$4.5$ & $-14.01030i$ & $-11.52580i$ &  $-10.57970i$ & $-9.70835i$ & $-8.81550i$ \\ \hline
\end{tabular}%
\end{table}

\section{Concluding comments}
\label{conclusion}
In this work we have calculated numerically the QNFs of scalar field perturbations of nonlinear charged Lifshitz black branes with hyperscaling violation by imposing suitable boundary conditions on the event horizon and at spatial infinity. The scalar field is considered as a mere test field, without backreaction over the spacetime itself. In this study we have considered the case $\alpha>-1$, additionally, for $\alpha>0$ we have restricted to massless scalar field $m=0$. Then, we have studied the stability of these black branes under massive and massless scalar field perturbations. In general, our results show that the QNFs have a negative imaginary part. Therefore, the black brane is stable under massive and massless scalar field perturbations. Also, we can see that there is a limit on the dynamical exponent $z$ above which the system is always overdamped for a given dimension, and the hyperscaling violating exponent shifts this limit. For instance, as we can observe in Table \ref{QNM3ml}, for $D=5$ and $\theta=-1$ we have found  that the system is non-overdamped for $z=3.5$, but it is overdamped for $z=4$. However,  for $D=5$ and $\theta=0$ we have found that the system is non-overdamped for $z=2.5$ and it is overdamped for $z=3$, and for $D=5$ and $\theta=0.5$ we have found that the system is overdamped for $z=2.5$. It is worth to mention that the shift also depends on the dimension. On the other hand, when we increase the hyperscaling violating exponent the relaxation time of the dual thermal states increases (due to the absolute value of the imaginary part of the QNFs decreases) and when we increase the nonlinear exponent the relaxation time increases too, when the system is overdamped.

\acknowledgments 

This work was funded by Comisi\'{o}n
Nacional de Ciencias y Tecnolog\'{i}a through FONDECYT Grant 11140674 (PAG). P. A. G. acknowledges the hospitality of the Universidad de La Serena where part of this work was undertaken.

\appendix


\begin{thebibliography}{99}

\bibitem{Kachru:2008yh} S.~Kachru, X.~Liu and M.~Mulligan, 
Phys.\ Rev.\ D \textbf{78}, 106005 (2008) [arXiv:0808.1725 [hep-th]]. 

\bibitem{Hartnoll:2009ns} 
  S.~A.~Hartnoll, J.~Polchinski, E.~Silverstein and D.~Tong,
  JHEP {\bf 1004}, 120 (2010)
  [arXiv:0912.1061 [hep-th]].
  
  
\bibitem{Brynjolfsson:2009ct} 
  E.~J.~Brynjolfsson, U.~H.~Danielsson, L.~Thorlacius and T.~Zingg,
  J.\ Phys.\ A {\bf 43}, 065401 (2010)
  [arXiv:0908.2611 [hep-th]].
  
\bibitem{Sin:2009wi} 
  S.~J.~Sin, S.~S.~Xu and Y.~Zhou,
  Int.\ J.\ Mod.\ Phys.\ A {\bf 26}, 4617 (2011)
  [arXiv:0909.4857 [hep-th]].
 
\bibitem{Schaposnik:2012cr} 
  F.~A.~Schaposnik and G.~Tallarita,
  Phys.\ Lett.\ B {\bf 720}, 393 (2013)
  [arXiv:1210.8358 [hep-th]].
  
\bibitem{Momeni:2012tw} 
  D.~Momeni, R.~Myrzakulov, L.~Sebastiani and M.~R.~Setare,
  arXiv:1210.7965 [hep-th].
  
\bibitem{Bu:2012zzb} 
  Y.~Bu,
  Phys.\ Rev.\ D {\bf 86}, 046007 (2012)
  [arXiv:1211.0037 [hep-th]].

    
  
\bibitem{Keranen:2012mx} 
  V.~Keranen and L.~Thorlacius,
  Class.\ Quant.\ Grav.\  {\bf 29}, 194009 (2012)
  [arXiv:1204.0360 [hep-th]].


\bibitem{Zhao:2013pva} 
  Z.~Zhao, Q.~Pan and J.~Jing,
  Phys.\ Lett.\ B {\bf 735}, 438 (2014)
  [arXiv:1311.6260 [hep-th]].
  
\bibitem{Lu:2013tza} 
  J.~W.~Lu, Y.~B.~Wu, P.~Qian, Y.~Y.~Zhao and X.~Zhang,
  Nucl.\ Phys.\ B {\bf 887}, 112 (2014)
  [arXiv:1311.2699 [hep-th]].
  
  
\bibitem{Tallarita:2014bga} 
  G.~Tallarita,
  Phys.\ Rev.\ D {\bf 89}, no. 10, 106005 (2014)
  [arXiv:1402.4691 [hep-th]].
  

\bibitem{Horowitz:2011gh} G.~T.~Horowitz and B.~Way, 
Phys.\ Rev.\ D \textbf{85} (2012) 046008 [arXiv:1111.1243 [hep-th]]. 


\bibitem{AyonBeato:2009nh} E.~Ayon-Beato, A.~Garbarz, G.~Giribet and
M.~Hassaine, 
Phys.\ Rev.\ D \textbf{80} (2009) 104029 [arXiv:0909.1347 [hep-th]]. 


\bibitem{Cai:2009ac} R.~-G.~Cai, Y.~Liu and Y.~-W.~Sun, 
JHEP \textbf{0910} (2009) 080 [arXiv:0909.2807 [hep-th]]. 


\bibitem{AyonBeato:2010tm} E.~Ayon-Beato, A.~Garbarz, G.~Giribet and
M.~Hassaine, 
JHEP \textbf{1004} (2010) 030 [arXiv:1001.2361 [hep-th]]. 


\bibitem{Dehghani:2010kd} M.~H.~Dehghani and R.~B.~Mann, 
JHEP \textbf{1007} (2010) 019 [arXiv:1004.4397 [hep-th]]. 


\bibitem{Mann:2009yx} R.~B.~Mann, 
JHEP \textbf{0906} (2009) 075 [arXiv:0905.1136 [hep-th]]. 

\bibitem{Balasubramanian:2009rx}
  K.~Balasubramanian and J.~McGreevy,
  Phys.\ Rev.\ D {\bf 80} (2009) 104039
  [arXiv:0909.0263 [hep-th]].


\bibitem{Bertoldi:2009vn} G.~Bertoldi, B.~A.~Burrington and A.~Peet, 
Phys.\ Rev.\ D \textbf{80} (2009) 126003 [arXiv:0905.3183 [hep-th]]. 


    



  
\bibitem{Devecioglu:2010sf} D.~O.~Devecioglu and O.~Sarioglu, 
Phys.\ Rev.\ D \textbf{83} (2011) 021503 [arXiv:1010.1711 [hep-th]]. 


\bibitem{Devecioglu:2011yi} D.~O.~Devecioglu and O.~Sarioglu, 
Phys.\ Rev.\ D \textbf{83} (2011) 124041 [arXiv:1103.1993 [hep-th]]. 


  \bibitem{Gim:2014nba} 
  Y.~Gim, W.~Kim and S.~H.~Yi,
  JHEP {\bf 1407}, 002 (2014)
  [arXiv:1403.4704 [hep-th]].


\bibitem{Gonzalez:2011nz} H.~A.~Gonzalez, D.~Tempo and R.~Troncoso, 
JHEP \textbf{1111} (2011) 066 [arXiv:1107.3647 [hep-th]]. 


\bibitem{Myung:2012cb} Y.~S.~Myung and T.~Moon, 
Phys.\ Rev.\ D \textbf{86} (2012) 024006 [arXiv:1204.2116 [hep-th]]. 


\bibitem{Myung:2012xc} Y.~S.~Myung, 
Eur.\ Phys.\ J.\ C \textbf{72} (2012) 2116 [arXiv:1203.1367 [hep-th]]. 

 
 \bibitem{Dong:2012se} 
  X.~Dong, S.~Harrison, S.~Kachru, G.~Torroba and H.~Wang,
  JHEP {\bf 1206}, 041 (2012)
  [arXiv:1201.1905 [hep-th]].  
  
  \bibitem{Gath:2012pg} 
  J.~Gath, J.~Hartong, R.~Monteiro and N.~A.~Obers,
  JHEP {\bf 1304}, 159 (2013)
  [arXiv:1212.3263 [hep-th]].
  
  \bibitem{Feng:2015yja}
  X.~H.~Feng and W.~J.~Geng,
  Phys.\ Lett.\ B {\bf 747}, 395 (2015)
  [arXiv:1502.00863 [hep-th]].

\bibitem{Dehghani:2015gza} 
  M.~H.~Dehghani, A.~Sheykhi and S.~E.~Sadati,
  Phys.\ Rev.\ D {\bf 91}, no. 12, 124073 (2015)
  [arXiv:1505.01134 [hep-th]].

\bibitem{Ganjali:2015cba} 
  M.~A.~Ganjali,
  arXiv:1508.05614 [hep-th].

  
  

  




\bibitem{Regge:1957td} T.~Regge and J.~A.~Wheeler, 
Phys.\ Rev.\ \textbf{108}, 1063 (1957).


\bibitem{Zerilli:1971wd} F.~J.~Zerilli, 
Phys.\ Rev.\ D \textbf{2}, 2141 (1970).


\bibitem{Zerilli:1970se} F.~J.~Zerilli, 
Phys.\ Rev.\ Lett.\ \textbf{24}, 737 (1970). 


\bibitem{Kokkotas:1999bd} K.~D.~Kokkotas and B.~G.~Schmidt, 
Living Rev.\ Rel.\ \textbf{2}, 2 (1999) [gr-qc/9909058]. 


\bibitem{Nollert:1999ji} H.~-P.~Nollert, 
Class.\ Quant.\ Grav.\ \textbf{16}, R159 (1999). 


\bibitem{Konoplya:2011qq} R.~A.~Konoplya and A.~Zhidenko, 
Rev.\ Mod.\ Phys.\ \textbf{83}, 793 (2011) [arXiv:1102.4014 [gr-qc]].


\bibitem{Maldacena:1997re} J.~M.~Maldacena, 
Adv.\ Theor.\ Math.\ Phys.\ \textbf{2}, 231 (1998) [hep-th/9711200]. 


\bibitem{Birmingham:2001pj} D.~Birmingham, I.~Sachs and S.~N.~Solodukhin, 
Phys.\ Rev.\ Lett.\ \textbf{88}, 151301 (2002) [hep-th/0112055].


\bibitem{Alishahiha:2012nm}  M.~Alishahiha, M.~R.~Mohammadi Mozaffar and
A.~Mollabashi,  
Phys.\ Rev.\ D \textbf{86} (2012) 026002  [arXiv:1201.1764 [hep-th]].  

\bibitem{Catalan:2013eza}
  M.~Catalan, E.~Cisternas, P.~A.~Gonzalez and Y.~Vasquez,
  Eur.\ Phys.\ J.\ C {\bf 74} (2014) 3,  2813
  [arXiv:1312.6451 [gr-qc]].


\bibitem{CuadrosMelgar:2011up} B.~Cuadros-Melgar, J.~de Oliveira and
C.~E.~Pellicer, 
Phys.\ Rev.\ D \textbf{85}, 024014 (2012) [arXiv:1110.4856 [hep-th]]. 


\bibitem{Gonzalez:2012de} P.~A.~Gonzalez, J.~Saavedra and Y.~Vasquez, 
Int.\ J.\ Mod.\ Phys.\ D \textbf{21}, 1250054 (2012) [arXiv:1201.4521
[gr-qc]]. 


\bibitem{Gonzalez:2012xc} P.~A.~Gonzalez, F.~Moncada and Y.~Vasquez, 
Eur.\ Phys.\ J.\ C \textbf{72}, 2255 (2012) [arXiv:1205.0582 [gr-qc]]. 


\bibitem{Becar:2012bj} R.~Becar, P.~A.~Gonzalez and Y.~Vasquez, 
Int.\ J.\ Mod.\ Phys.\ D \textbf{22}, 1350007 (2013) [arXiv:1210.7561
[gr-qc]]. 


\bibitem{Giacomini:2012hg} A.~Giacomini, G.~Giribet, M.~Leston, J.~Oliva and
S.~Ray, 
Phys.\ Rev.\ D \textbf{85} (2012) 124001 [arXiv:1203.0582 [hep-th]]. 

  

\bibitem{Lepe:2012zf}
  S.~Lepe, J.~Lorca, F.~Pena and Y.~Vasquez,
  Phys.\ Rev.\ D {\bf 86} (2012) 066008
  [arXiv:1205.4460 [hep-th]].
  
  \bibitem{Catalan:2014ama} 
  M.~Catalan, E.~Cisternas, P.~A.~Gonzalez and Y.~Vasquez,
  arXiv:1404.3172 [gr-qc].
  
  \bibitem{Catalan:2014una} 
  M.~Catal\'{a}n and Y.~V\'{a}squez,
  Phys.\ Rev.\ D {\bf 90}, no. 10, 104002 (2014)
  [arXiv:1407.6394 [gr-qc]].
  
  
  
  
  
  
  
 
   \bibitem{Lopez-Ortega:2014oha} 
  A.~L�pez-Ortega,
  Gen.\ Rel.\ Grav.\  {\bf 46}, 1756 (2014)
  [arXiv:1406.0126 [gr-qc]]. 
  
  
  
  



\bibitem{Sybesma:2015oha} 
  W.~Sybesma and S.~Vandoren,
  JHEP {\bf 1505}, 021 (2015)
  [arXiv:1503.07457 [hep-th]].


\bibitem{Cho:2009cj} H.~T.~Cho, A.~S.~Cornell, J.~Doukas and W.~Naylor,
Class.\ Quant.\ Grav.\ \textbf{27} (2010) 155004 [arXiv:0912.2740 [gr-qc]]. 


\bibitem{Ciftci}
H. Ciftci, R. L. Hall, and N. Saad, 
J. Phys. A \textbf{36}(47), 11807-11816 (2003).

\bibitem{Ciftci:2005xn}
  H.~Ciftci, R.~L.~Hall and N.~Saad,
  Phys.\ Lett.\ A {\bf 340} (2005) 388.
  

\bibitem{Cho:2011sf} H.~T.~Cho, A.~S.~Cornell, J.~Doukas, T.~R.~Huang and
W.~Naylor, 
Adv.\ Math.\ Phys.\ \textbf{2012} (2012)
281705 [arXiv:1111.5024 [gr-qc]]. 


\bibitem{Zhang:2015jda}
  C.~Y.~Zhang, S.~J.~Zhang and B.~Wang,
  arXiv:1501.03260 [hep-th].
  
  


\bibitem{Barakat:2006ki} T.~Barakat, 
Int.\ J.\ Mod.\ Phys.\ A \textbf{21} (2006) 4127. 

  
  \bibitem{Horowitz:1999jd} 
  G.~T.~Horowitz and V.~E.~Hubeny,
  Phys.\ Rev.\ D {\bf 62}, 024027 (2000)
  [hep-th/9909056].






      
\end{thebibliography}
\end{document}